\documentclass[11pt]{article}
\usepackage{amsmath,epsfig,sint}
\font\sixrm=cmr6

\def\rmD{{\rm D}}
\def\rme{{\rm e}}
\def\rmO{{\rm O}}

\def\bfp{{\bf p}}
\def\bfq{{\bf q}}

\def\bfx{{\bf x}}
\def\bfy{{\bf y}}
\def\bfz{{\bf z}}

\def\proof{\noindent{\sl Proof:}\kern0.6em}

\def\frac#1#2{\hbox{$#1\over#2$}}
\def\dual{\mathstrut^*\kern-0.1em}

\def\ring{\mathaccent"7017}
\def\lvec#1{\setbox0=\hbox{$#1$}
    \setbox1=\hbox{$\scriptstyle\leftarrow$}
    #1\kern-\wd0\smash{
    \raise\ht0\hbox{$\raise1pt\hbox{$\scriptstyle\leftarrow$}$}}
    \kern-\wd1\kern\wd0}
\def\rvec#1{\setbox0=\hbox{$#1$}
    \setbox1=\hbox{$\scriptstyle\rightarrow$}
    #1\kern-\wd0\smash{
    \raise\ht0\hbox{$\raise1pt\hbox{$\scriptstyle\rightarrow$}$}}
    \kern-\wd1\kern\wd0}

\def\nab#1{{\nabla_{#1}}}
\def\nabstar#1{\nabla\kern-0.5pt\smash{\raise 4.5pt\hbox{$\ast$}}
               \kern-4.5pt_{#1}}

\def\drvstar#1{\partial\kern-0.5pt\smash{\raise 4.5pt\hbox{$\ast$}}
               \kern-5.0pt_{#1}}

\def\momp#1#2{
    \setbox0=\hbox{${#1}$}\setbox1=\hbox{${#1}_{#2}$}
    {#1}_{#2}\kern-\wd1\kern\wd0
    \smash{\raise4.5pt\hbox{$\scriptscriptstyle +$}}}
\def\momm#1#2{
    \setbox0=\hbox{${#1}$}\setbox1=\hbox{${#1}_{#2}$}
    {#1}_{#2}\kern-\wd1\kern\wd0
    \smash{\raise4.5pt\hbox{$\scriptscriptstyle -$}}}
\def\mompm#1#2{
    \setbox0=\hbox{${#1}$}\setbox1=\hbox{${#1}_{#2}$}
    {#1}_{#2}\kern-\wd1\kern\wd0
    \smash{\raise4.5pt\hbox{$\scriptscriptstyle \pm$}}}
\def\smomp#1#2{
    \setbox0=\hbox{${#1}$}\setbox1=\hbox{${#1}_{#2}$}
    {#1}_{#2}\kern-\wd1\kern\wd0
    \smash{\raise3pt\hbox{$\scriptscriptstyle +$}}}
\def\smomm#1#2{
    \setbox0=\hbox{${#1}$}\setbox1=\hbox{${#1}_{#2}$}
    {#1}_{#2}\kern-\wd1\kern\wd0
    \smash{\raise3pt\hbox{$\scriptscriptstyle -$}}}
\def\smompm#1#2{
    \setbox0=\hbox{${#1}$}\setbox1=\hbox{${#1}_{#2}$}
    {#1}_{#2}\kern-\wd1\kern\wd0
    \smash{\raise3pt\hbox{$\scriptscriptstyle \pm$}}}
\def\si{\kern1pt{\rm si}}
\def\co{\kern1pt{\rm co}}

\def\Nf{N_{\rm f}}
\def\psibar{\bar{\psi}}

\def\psiclass{\psi_{\rm cl}}
\def\psibarclass{\psibar_{\rm cl}}

\def\psiprime{\psi\kern1pt'}
\def\psibarprime{\psibar\kern1pt'}
\def\rhoprime{\rho\kern1pt'}
\def\rhobar{\bar{\rho}}
\def\rhobarprime{\rhobar\kern1pt'}
\def\rhobartilde{\kern2pt\tilde{\kern-2pt\rhobar}}
\def\rhobartildeprime{\kern2pt\tilde{\kern-2pt\rhobar}\kern1pt'}

\def\zetabar{\bar{\zeta}}
\def\zetaprime{\zeta\kern1pt'}
\def\zetabarprime{\zetabar\kern1pt'}
\def\zetar{\zeta_{\raise-1pt\hbox{\sixrm R}}}
\def\zetabarr{\zetabar_{\raise-1pt\hbox{\sixrm R}}}

\def\phiimpr{\phi_{\kern0.5pt\hbox{\sixrm I}}}

\def\dirac#1{\gamma_{#1}}
\def\diracstar#1#2{
    \setbox0=\hbox{$\gamma$}\setbox1=\hbox{$\gamma_{#1}$}
    \gamma_{#1}\kern-\wd1\kern\wd0
    \smash{\raise4.5pt\hbox{$\scriptstyle#2$}}}

\def\ba{b_{\rm A}}
\def\tba{\tilde{b}_{\rm A}}
\def\bp{b_{\rm P}}

\def\bv{b_{\rm V}}
\def\tbv{\tilde{b}_{\rm V}}

\def\bg{b_{\rm g}}
\def\bm{b_{\rm m}}
\def\tbm{\tilde{b}_{\rm m}}
\def\bmu{b_{\mu}}
\def\bzeta{b_{\zeta}}

\def\tbvbar{\tilde{b}_{\rm \bar{V}}}
\def\cvbar{c_{\rm \bar{V}}}

\def\ca{c_{\rm A}}
\def\cv{c_{\rm V}}
\def\Ca{\ca}

\def\csw{c_{\rm sw}}

\def\ctildes{\tilde{c}_{\rm s}}
\def\ctildet{\tilde{c}_{\rm t}}

\def\fa{f_{\rm A}}

\def\fp{f_{\rm P}}
\def\fx{f_{\rm X}}

\def\fv{f_{\rm V}}

\def\f1{f_1}

\def\ha{h_{\rm A}}
\def\hda{h_{\rm dA}}
\def\hdv{h_{\rm dV}}
\def\hp{h_{\rm P}}
\def\hv{h_{\rm V}}
\def\h1{h_1}

\def\tr{\,\hbox{tr}\,}

\def\CF{C_{\rm F}}
\def\cf{\CF}

\def\Sg{S_{\rm G}}
\def\Sf{S_{\rm F}}
\def\Seff{S_{\rm eff}}
\def\Simpr{S_{\rm impr}}

\def\opprime#1{\setbox0=\hbox{${\cal O}$}\setbox1=\hbox{${\cal O}_{\rm #1}$}
    {\cal O}_{\rm #1}\kern-\wd1\kern\wd0
    \smash{\raise4.5pt\hbox{\kern1pt$\scriptstyle\prime$}}\kern1pt}
\def\ophat#1{\widehat{\cal O}_{\rm #1}}
\def\ophatprime#1{\setbox0=\hbox{$\widehat{\cal O}$}
    \setbox1=\hbox{$\widehat{\cal O}_{\rm #1}$}
    \widehat{\cal O}_{\rm #1}\kern-\wd1\kern\wd0
    \smash{\raise4.5pt\hbox{\kern1pt$\scriptstyle\prime$}}\kern1pt}

\def\bopprime#1{\setbox0=\hbox{${\cal O}$}\setbox1=\hbox{${\cal O}_{\rm #1}$}
    {\cal L}_{\rm #1}\kern-\wd1\kern\wd0
    \smash{\raise4.5pt\hbox{\kern1pt$\scriptstyle\prime$}}\kern1pt}

\def\blagprime#1{\setbox0=\hbox{${\cal B}$}\setbox1=\hbox{${\cal B}_{#1}$}
    {\cal B}_{#1}\kern-\wd1\kern\wd0
    \smash{\raise5.2pt\hbox{\kern1pt$\scriptstyle\prime$}}\kern1pt}

\def\gr{g_{{\hbox{\sixrm R}}}}

\def\muq{\mu_{\rm q}}
\def\mq{m_{\rm q}}
\def\mqtilde{\widetilde{m}_{\rm q}}
\def\muqtilde{\widetilde{\mu}_{\rm q}}

\def\mr{m_{{\hbox{\sixrm R}}}}
\def\mur{\mu_{{\hbox{\sixrm R}}}}
\def\mc{m_{\rm c}}
\def\mpole{m_{\rm p}}

\def\za{Z_{\rm A}}
\def\zv{Z_{\rm V}}
\def\zp{Z_{\rm P}}

\def\zg{Z_{\rm g}}
\def\zm{Z_{\rm m}}

\def\zzeta{Z_{\zeta}}

\def\Za{\za}
\def\Zv{\zv}
\def\Zp{\zp}

\def\Zg{\zg}
\def\Zm{\zm}
\def\Zmu{Z_{\mu}}

\def\Zz{\zzeta}

\def\gtilde{\tilde{g}_0}

\def\msbar{{\rm \overline{MS\kern-0.05em}\kern0.05em}}

\newcommand{\bes}{\begin{eqnarray}}
\newcommand{\ees}{\end{eqnarray}}

\begin{document}
\begin{titlepage}
\begin{flushright}
   CERN-TH/2001-014\\
   MPI-PhT/2000-52\\
   Bicocca-FT-01-02\\
   April 2001
\end{flushright}
\vskip 0.5 cm
\begin{center}
  {\Large\bf O($a$) improved twisted mass lattice QCD \\[1.5ex]
   }
\end{center}
\vskip 0.5 cm
\begin{figure}[h]
\begin{center}
\epsfig{figure=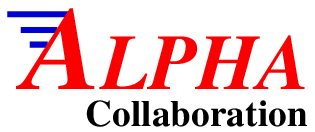} 
\end{center}
\end{figure}
\begin{center}
{\large Roberto Frezzotti$^{\scriptscriptstyle a,b}$,
        Stefan Sint$^{\scriptscriptstyle b}$
    and Peter Weisz$^{\scriptscriptstyle c}$}
\end{center}
\vskip 2.3ex
\begin{flushleft}
$^{\scriptstyle a}$ Universit\`a di Milano-Bicocca, 
Dipartimento di Fisica, Piazza della Scienza~3,
I--20126 Milano, Italy\\[1ex]
$^{\scriptstyle b}$ CERN, Theory Division,
CH--1211 Geneva 23, Switzerland\\[1ex]
$^{\scriptstyle c}$ Max-Planck-Institut f\"ur Physik,
F\"ohringer Ring 6, D--80805 M\"unchen, Germany
\end{flushleft}
\begin{center} 
 {\bf Abstract}
\end{center}
\vskip 0.7ex
Lattice QCD with Wilson quarks and a chirally twisted mass 
term (tmQCD) has been introduced in refs.~\cite{lat99,tmQCD1}.
We here apply Symanzik's improvement programme to this theory
and list the counterterms which arise at first order in the
lattice spacing $a$. Based on the generalised transfer matrix, 
we define the tmQCD Schr\"odinger functional and use it
to derive renormalized on-shell correlation functions.
By studying their continuum approach in perturbation theory
we then determine the new O($a$) counterterms of the action 
and of a few quark bilinear operators to one-loop order.

\vfill
\eject

\end{titlepage}

\section{Introduction}

In ref.~\cite{tmQCD1} twisted mass 
lattice QCD (tmQCD) has been introduced as a solution
to the problem of unphysical fermion zero modes which
plague standard lattice QCD with quarks of the Wilson type.
We will assume that the reader is familiar with the
motivation of this approach, and refer to~\cite{lat99}
for an introduction. The main topic of the 
present paper is the application of Symanizik's 
improvement programme to tmQCD. We introduce the set-up
in the simplest case of two mass-degenerate quarks,
and study the improved action and the improved 
composite fields which appear in the PCAC and PCVC relations. 

Our strategy follows closely refs.~\cite{paperI,paperII,PeterStefan}:
in section~2 we go through the structure of the O($a$) improved theory.
We then define the Schr\"odinger functional for tmQCD, 
and use it to derive suitable on-shell correlation functions (sect.~3).
The perturbation expansion is then carried out along the lines
of ref.~\cite{PeterStefan}, and the new O($a$) improvement 
coefficients are obtained at the tree-level in sect.~4 and
to one-loop order in sect.~5.
A few details have been delegated to appendices.
Appendix~A describes how the twisted mass term can be incorporated
in L\"uscher's construction of the transfer matrix~\cite{transfer}, 
and Appendix~B contains the analytic expressions
for the coefficients used in the analysis 
of the one-loop calculation.

\section{Renormalized and O($a$) improved tmQCD}

The renormalization procedure for twisted mass lattice QCD
with Wilson quarks has already been discussed in ref.~\cite{tmQCD1}.
Here we apply Symanzik's improvement programme
to first order in the lattice spacing
$a$. The procedure is standard and the details of its application to lattice
QCD with $\Nf$ mass degenerate Wilson quarks can 
be found in ref.~\cite{paperI}.

Our starting point is the unimproved tmQCD lattice action
for a doublet of mass degenerate quarks,
\begin{equation}
  S[U,\psibar,\psi]=\Sg[U]+\Sf[U,\psibar,\psi], 
\end{equation}
with the standard Wilson gauge action and the fermionic part
\begin{equation}
  \Sf[U,\psibar,\psi] = a^4\sum_{x}
  \psibar(x)\left(D+m_0+i\muq\gamma_5\tau^3\right)\psi(x).
  \label{tmLQCD}
\end{equation}
The massless Wilson-Dirac operator is given by
\begin{equation}
  D=\frac12\sum_{\mu}\Bigl\{
       \left(\nab\mu+\nabstar\mu\right)\gamma_\mu - a\nabstar\mu\nab\mu  
                     \Bigr\}, 
\end{equation}
where the forward and backward covariant lattice derivatives in direction
$\mu$ are denoted by $\nab\mu$ and $\nabstar\mu$ respectively.
As tmQCD with vanishing twisted
mass parameter $\muq$ reduces to standard lattice QCD
we expect that improvement is achieved
by using the standard O($a$) improved
theory and adding the appropriate O($a$) counterterms which
are proportional to (powers of) $\muq$,
and which are allowed by the lattice symmetries.
The procedure hence consists in a straightforward extension
of ref.~\cite{paperI}, and we take over notation and conventions from 
this reference without further notice.

\subsection{Renormalized O($a$) improved parameters}

Following ref.~\cite{paperI} we assume that a 
mass-independent renormalization scheme has been chosen,
and we take the same steps as done there 
for standard lattice QCD. At $\muq=0$ the Sheikholeslami-Wohlert 
term~\cite{SW} suffices to improve the action, 
up to a rescaling of the bare parameters by terms proportional to the 
subtracted bare mass $\mq=m_0-\mc$~\cite{paperI}. 
At non-vanishing $\muq$ we find that improved bare
parameters are of the form
\begin{eqnarray}
     \gtilde^2 &=& g_0^2(1+\bg a\mq),\\
   \mqtilde    &=& \mq(1+ b_{\rm m} a\mq) + \tbm a\muq^2,\\
   \muqtilde   &=& \muq(1+\bmu a\mq),
\end{eqnarray}
i.e.~there exist two new counterterms with coefficients
$\bmu$ and $\tbm$. 
The renormalized O($a$) improved mass and coupling
constant are then proportional to these parameters, viz.
\begin{eqnarray}
  \gr^2 &=&  \gtilde^2\Zg(\gtilde^2,a\mu), \\ 
  \mr   &=&  \mqtilde\Zm(\gtilde^2,a\mu),  \\
  \mur  &=&  \muqtilde Z_\mu(\gtilde^2,a\mu).
\end{eqnarray}
The ratio of the appropriately renormalized mass parameters
determines the angle $\alpha$ which is involved in the 
physical interpretation of the theory~\cite{tmQCD1}. We will discuss
below the general O($a$) improved definition of $\alpha$. Here
we note that the case of particular interest, $\alpha=\pi/2$, corresponds
to $\mr=0$, which implies $\mq=\rmO(a)$~\cite{tmQCD1}. 
In this case all the usual $b$-coefficients multiply terms of 
O($a^2$) and are thus negligible in the spirit of O($a$) improvement. 
One then remains with a single coefficient $\tbm$, which compares favorably
to the situation in standard lattice QCD where 
two coefficients, $\bm$ and $\bg$, are required. 

\subsection{Renormalized O($a$) improved composite fields}

We assume that composite fields are renormalized in 
a mass-independent scheme, and such that the tmQCD Ward identities
are respected~\cite{tmQCD1}. Attention will be restricted to 
the quark bilinear operators which appear in the PCAC and
PCVC relations. Moreover, we only consider the first two 
flavour components, and thus avoid the
renormalization of power divergent operators 
such as the iso-singlet scalar density~\cite{tmQCD1}. 
As explained in ref.~\cite{tmQCD1}, the third flavour component
of the PCAC and PCVC relations can be inferred in the continuum limit,
by assuming the restoration of the physical isospin symmetry.
The O($a$) improved currents and pseudo-scalar 
density with indices $a,b \in \{1,2\}$ are then parameterised as follows,
\begin{eqnarray}
  (A_{\rm R})_\mu^a &=& \Za(1+\ba a\mq)
                          \Bigl[A_\mu^a+\ca a\tilde\partial_\mu P^a + a\muq\,
                          \tba\,
                          \varepsilon^{3ab} V_\mu^b \Bigr], \\
  (V_{\rm R})_\mu^a &=& \Zv(1+\bv a\mq)
                          \Bigl[V_\mu^a+\cv 
                          a\tilde\partial_\nu T^a_{\mu\nu}+ 
                          a\muq\,\tbv\,
                          \varepsilon^{3ab} A_\mu^b \Bigr],\label{locvec}\\
  (P_{\rm R})^a     &=& \Zp(1+\bp a\mq)\,P^a.  \label{axial}
\end{eqnarray}
Here we have chosen the bare operators which are  
local on the lattice, with the conventions of 
ref.~\cite{paperI}. While this is the simplest choice, 
we also recall the definition of the point-split vector current,
\begin{eqnarray}
   \widetilde{V}^a_\mu(x) &=&  
 \frac12 \Bigl\{\psibar(x)(\gamma_\mu-1)\frac{\tau^a}{2}
 U(x,\mu)\psi(x+a\hat\mu) \nonumber\\
 && \hphantom{\frac12 } +\psibar(x+a\hat\mu)(\gamma_\mu+1)\frac{\tau^a}{2} 
 U(x,\mu)^{-1}\psi(x)\Bigr\},
 \label{pointsplit}
\end{eqnarray}
which is obtained  through a vector variation of the action.
This current is protected against renormalization,
and the PCVC relation
\begin{equation}
    \partial^{\ast}_\mu \tilde{V}^a_\mu(x) = -2\muq\varepsilon^{3ab} P^b(x),
   \label{PCVC}
\end{equation}
is an {\em exact} lattice identity, with the local
pseudo-scalar density and the backward derivative 
$\partial^{\ast}_\mu$ in $\mu$-direction~\cite{tmQCD1}. 
This implies the identity $Z_\mu\Zp=1$ in any
renormalization scheme which respects the PCVC relation.

\subsection{An alternative definition of the improved vector current}

An alternative renormalized improved current
can be obtained from the point-split current~(\ref{pointsplit}). 
For this it is convenient to start from the symmetrized version
\begin{equation}
   {\bar V}_\mu^a(x)=  \frac12\left(\widetilde{V}^a_\mu(x)+  
                            \widetilde{V}^a_\mu(x-a\hat\mu)\right),
\end{equation}
which behaves under space-time reflections in the same way
as the local vector current. The counterterm structure then
is the same as in eq.~(\ref{locvec}), i.e.~one finds
\begin{equation}
  ({\bar V}_{\rm R})^a_\mu = Z_{\bar{\rm V}}(1+b_{\bar{\rm V}}a\mq)
  \left[{\bar V}_\mu^a + 
   \cvbar a\tilde\partial_\nu T^a_{\mu\nu}
    +\tbvbar a\muq\varepsilon^{3ab}A^b_\mu\right], 
  \label{renvec}
\end{equation}
where we have again restricted the indices $a,b$ to the
first two components.  One may now easily show that 
\begin{equation}
  Z_{\bar{\rm V}}=1,\qquad b_{\bar{\rm V}}=0.
 \label{mult_ren}
\end{equation}
To see this we first note that at $\muq=0$ the vector charge 
of this current is given by
\begin{equation}
 {Q}^a_{\bar{\rm V}}(t) = \frac12 Z_{\bar{\rm V}}(1+b_{\bar{\rm V}}a\mq)
 \left[Q^a_{\rm V}(t)+Q^a_{\rm V}(t-a)\right],
  \label{charge_impr}
\end{equation}
with
\begin{equation}
  Q_{\rm V}^a(x_0)=a^3\sum_{\bfx} \widetilde{V}^a_0(x).
\end{equation}
At $\muq=0$, correlation functions of the charge 
are $x_0$-independent\footnote{i.e.~as 
long as the time ordering of the space-time arguments 
in the given correlation function remains unchanged.},
and the O($a$) improved charge algebra for ${Q}^a_{\bar{\rm V}}$ and the
exact charge algebra for $Q^a_{\rm V}$ together imply
that the whole renormalization factor in eq.~(\ref{charge_impr}) must be 
unity. As this holds independently of $\mq$, 
one arrives at the conclusion~(\ref{mult_ren}).

A further relation is obtained by noting that the 
PCVC relation between the renormalized O($a$) improved fields,
\begin{equation}
  \tilde\partial_\mu({\bar V}_{\rm R})^a_\mu 
  =-2\mur\varepsilon^{3ab}(P_{\rm R})^b, 
\end{equation}
with the symmetric derivative $\tilde\partial_\mu
=\frac12(\partial_\mu+\partial_\mu^\ast)$ 
must hold up to O($a^2$) corrections. 
Then, using the identity
\begin{equation}
   \tilde\partial_\mu  {\bar V}^a_\mu(x) = 
   \partial_\mu^\ast\left(\widetilde{V}^a_\mu(x)
   +\frac14 a^2\partial_\mu^\ast\partial_\mu^{} \widetilde{V}^a_\mu(x)\right), 
  \label{sym_div}
\end{equation}
one obtains the relation
\begin{equation}
  Z_{\rm P}\Zm\Za^{-1}\tbvbar=-(\bmu+\bp).
  \label{rel_impr}
\end{equation}
The scale-independent combination of renormalization constants
multiplying $\tbvbar$ is determined by axial 
Ward identities~\cite{babp}, so that eq.~(\ref{rel_impr}) can
be considered a relation between improvement coefficients.

\subsection{O($a$) improved definition of the angle $\alpha$}

The physical interpretation of the correlation functions 
in tmQCD depends on the angle $\alpha$, which is defined through
\begin{equation}
  \tan\alpha={\mur\over\mr}.
\end{equation}
In this equation $\mur$ and $\mr$ are the O($a$) improved renormalized mass
parameters which appear in the PCAC and  PCVC relations~\cite{tmQCD1}.
Up to terms of O($a^2$) we then find
\begin{equation}
  {\mur\over\mr} =  
 {{\muq[1+(b_\mu-\bm)a\mq]}\over{\Zp\Zm[\mq+\tbm a\muq^2]}}
  ={{\muq[1+(b_\mu+\bp-\ba)a\mq]}\over{\Za[m+\tba a\muq^2\Zv^{-1}]}}.
 \label{massratio}
\end{equation}
Here, $m$ denotes a bare mass which is obtained from some
matrix element of the PCAC relation involving  the unrenormalized
axial current $A^1_\mu +\ca\partial_\mu P^1$ and the local
density $P^1$. Given $m$, the critical mass $\mc$, and the finite
renormalization constants $\Za$, $\Zv$ and $\Zp\Zm$, 
the determination of the O($a$) improved angle 
requires the knowledge of two (combinations of)
improvement coefficients, which may be chosen to be $b_\mu-\bm$ and 
$\tbm$, or $b_\mu+\bp-\ba$ and $\tba$.
A special case is again $\alpha=\pi/2$, which is obtained for 
vanishing  denominators in eq.~(\ref{massratio}).
For this it is sufficient to know either $\tba$ or $\tbm$, and 
the finite renormalization constants $\Za$ or $\Zp\Zm$ are then not
needed.

\subsection{Redundancy of improvement coefficients}

Having introduced all O($a$) counterterms allowed by the
lattice symmetries, it is guaranteed that there exists a choice 
for the improvement coefficients such that 
O($a$) lattice artefacts in on-shell correlation functions 
are completely eliminated.
We now want to show that there is in fact a redundancy in the 
set of the new counterterms introduced so far, i.e.~the counterterms are
not unambiguously determined by the requirement of on-shell
improvement alone. To see this we consider the renormalized 
2-point functions 
\begin{eqnarray}
   G_{\rm A}(x-y) &=& \left\langle (A_{\rm R})^1_0(x)(P_{\rm R})^1(y)
   \right\rangle,\\
   G_{\rm V}(x-y) &=& \left\langle (V_{\rm R})^2_0(x)(P_{\rm R})^1(y)
  \right\rangle,
\end{eqnarray}
of the renormalized O($a$) improved fields defined in subsect.~2.2.
We assume that a quark mass independent
renormalization scheme has been chosen, and with the 
proper choice for the improvement coefficients one finds,
\begin{equation}
  G_{\rm X}(x) = \lim_{a\rightarrow 0}G_{\rm X}(x) + \rmO(a^2),
  \qquad {\rm X}={\rm A,V},
 \label{contapproach}
\end{equation}
provided that $x$ is kept non-zero in physical units.
If the new improvement coefficients
$\tbm,\,b_\mu,\,\tba$ and $\tbv$ 
were all necessary any change of O(1) in these coefficients would
introduce uncancelled O($a$) artefacts in eq.~(\ref{contapproach}).
Varying the coefficients $\tbm\rightarrow \tbm+\Delta\tbm$, 
$b_\mu \rightarrow b_\mu+\Delta b_\mu$ and $\tba \rightarrow
\tba+\Delta\tba$ in the correlation function $G_{\rm A}(x)$, we find 
that the correlation function itself changes
according to
\begin{eqnarray}
  \Delta G_{\rm A}(x) &=& -a\mur\Zp\Bigl[\Delta \tbm \Zp\Zm\,\mur 
  {\partial\over{\partial\mr}}G_{\rm A}(x) \nonumber\\
   && \hphantom{a}
   +\Delta b_\mu (\Zp\Zm)^{-1} \mr 
 {\partial\over{\partial\mur}}G_{\rm A}(x)  \nonumber\\
  && \hphantom{a}
  - \Delta \tba \Za \Zv^{-1} G_{\rm V}(x)\Bigr],
 \label{vary}
\end{eqnarray}
where terms of O($a^2$) have been neglected. In the derivation
of this equation one has to be careful to correctly take into account
the counterterms proportional to $b_\mu$ and $\tbm$. First of all 
we notice that changing an O($a$) counterterm can only 
induce changes of O($a$) in the correlation function.
For instance, the equation
\begin{equation}
 G_{\rm A}(x)\Bigr\vert_{b_\mu\rightarrow b_\mu+\Delta b_\mu} = 
 G_{\rm A}(x) + 
 \Delta b_\mu  {\partial\over{\partial{b_\mu}}} G_{\rm A}(x) + \rmO(a^2),
\end{equation}
holds even for finite changes $\Delta b_\mu$.
Second, when taking the continuum limit 
the bare mass parameters become functions
of the improvement coefficients such that the renormalized
O($a$) improved masses are fixed. For instance
one has
\begin{equation}
  \muq = \Zp\mur(1-b_\mu \Zm^{-1} a\mr) + \rmO(a^2),
  \label{tune_muq}
\end{equation}
and a straightforward application of the chain rule leads to
\begin{equation}
 {\partial\over{\partial{b_\mu}}} G_{\rm A}(x) = 
 \left({\partial \muq\over{\partial{b_\mu}}}\right)
 {\partial\over{\partial{\muq}}} G_{\rm A}(x) = -a\mur\mr\Zp \Zm^{-1}
 {\partial\over{\partial{\muq}}} G_{\rm A}(x),
\end{equation}
where we have used eq.~(\ref{tune_muq}) and neglected terms of O($a^2$).
Proceeding in the same way for the variation with respect to $\tbm$, 
and changing to renormalized  
parameters $\muq=\Zp\mur+\rmO(a)$, $\mq=Z_m^{-1}\mr +\rmO(a)$
eventually leads to eq.~(\ref{vary}).

At this point we recall eq.~(3.13) of ref.~\cite{tmQCD1},
which expresses the re-parameterization invariance 
with respect to changes of the angle $\alpha$. 
In terms of the above correlation functions
one finds, up to cutoff effects, 
\begin{equation}
  {\partial\over{\partial\alpha}}G_{\rm A}(x)\equiv
 \left(\mr{\partial\over{\partial\mur}}
      -\mur{\partial\over{\partial\mr}}\right)G_{\rm A}(x)
  = - G_{\rm V}(x).
\end{equation}
As a consequence not all the terms in eq.~(\ref{vary})
are independent, and the requirement that $\Delta G_{\rm A}(x)$ be of 
order $a^2$ entails only two conditions,
\begin{eqnarray}
  \Delta \tbm + \Delta b_\mu (\Zp\Zm)^{-2} &=&0, \\
  \Delta \tbm - \Delta \tba (\Zp\Zm \Zv)^{-1}\Za &=&0.
\end{eqnarray}
This makes precise the redundancy or over-completeness
of the counterterms alluded to above. The same procedure
applies to $G_{\rm V}(x)$, and we conclude that 
the requirement of on-shell O($a$) improvement only determines
the combinations of improvement coefficients 
$\tbm +  b_\mu (\Zp\Zm)^{-2}$, 
$\tbm - \tbv(\Zp\Zm \Za)^{-1}\Zv$, and
$\tbm - \tba(\Zp\Zm \Zv)^{-1}\Za$.
We emphasize that this redundancy is a generic feature of tmQCD,
and not linked to special choices for the fields or correlation
functions. In particular
we note that the third component of 
the axial variation of any composite field $\phi$ has the correct
quantum numbers to appear as an O($a\muq$) counterterm to $\phi$
itself.

In conclusion, O($a$) improved tmQCD as defined here constitutes
a one-parameter family of improved theories. 
In view of practical applications it is most convenient 
to  choose $\tbm$ as the free parameter and set it to some numerical value.
For reasons to become clear in section~4 our preferred choice is 
$\tbm=-\frac12$.
However, in the following we will keep all coefficients as unknowns
and only make a choice at the very end.
In order to define on-shell correlation functions which 
are readily accessible to perturbation theory we will first 
define the Schr\"odinger functional for tmQCD. It is then straightforward
to extend the techniques of refs.~\cite{paperII,PeterStefan}
to tmQCD and study the continuum approach of correlation functions 
derived from the Schr\"odinger functional.

\section{The Schr\"odinger functional for tmQCD}

This section follows closely Section~5 of ref.~\cite{paperI} and
ref.~\cite{paperII}. The reader will be assumed familiar with
these references, and we will refer to equations there
by using the prefix $\rm I$ and $\rm II$, respectively.

\subsection{Definition of the Schr\"odinger functional}

To define the Schr\"odinger functional for twisted mass lattice QCD,
it is convenient to follow refs.~\cite{LNWW,StefanI}.
The Schr\"odinger functional is thus obtained as the integral
kernel of some integer power $T/a$ of the transfer
matrix. Its Euclidean representation is given by
\begin{equation}
  {\cal Z}[\rhoprime,\rhobarprime,C'; \rho,\rhobar,C] = 
   \int \rmD[U]\rmD[\psi] \rmD[\psibar]\,\, \rme^{-S[U,\psibar,\psi]},
 \label{SF}
\end{equation}
and is thus considered as a functional of the fields at Euclidean
times $0$ and $T$. From the structure of the transfer matrix it follows 
that the boundary conditions for all fields are the same
as in the standard framework. In particular,
the quark fields satisfy,
\begin{equation} 
  \begin{split}
   {P_+\psi|}_{x_0=0} =  \rho,& \qquad 
   {P_-\psi|}_{x_0=T} =  \rhoprime, \\[1ex]       
   {\bar\psi P_-|}_{x_0=0}  =  \rhobar,& \qquad 
   {\bar\psi P_+|}_{x_0=T}   = \rhobarprime, \label{bcs} 
  \end{split}
\end{equation}
with the usual projectors $P_\pm=\frac12(1\pm\gamma_0)$.
The gauge field boundary conditions are as in eqs.(I.4.1)--(I.4.2)
and will not be repeated here.

The action in eq.~(\ref{SF}),
\begin{equation}
  S[U,\psibar,\psi] = \Sg[U]+\Sf[U,\psibar,\psi],
\end{equation}
splits into the gauge part~(I.4.5) and the quark action, which
assumes the same form as on the infinite lattice~(\ref{tmLQCD}). 
Note that we adopt the same conventions as in  
subsect.~4.2 of \cite{paperI}, in particular the quark and
antiquark fields are extended to all times by ``padding'' with
zeros, and the covariant derivatives 
in the finite space-time volume now contain
the additional phase factors related to $\theta_k$, ($k=1,2,3$).

\subsection{Renormalization and O($a$) improvement}

Renormalizability of the tmQCD Schr\"odinger functional 
could be verified along the lines of ref.~\cite{StefanII}.
However, this is not necessary as any new counterterm
is expected to be proportional to the twisted mass parameter
and is therefore at least of mass dimension~4.
One therefore expects the Schr\"odinger functional to be finite after
renormalization of the mass parameters and 
the gauge coupling as in infinite volume~\cite{tmQCD1},
and by scaling the  quark and anti-quark boundary fields
with a common renormalization constant~\cite{StefanII}.
This expectation will be confirmed in the course of the
perturbative calculation.

The structure of the new counterterms at O($a$) is again
determined by the symmetries. These are the same as 
in infinite space-time volume, 
except for those which exchange spatial and temporal directions.
The improved action,
\begin{equation}
  \Simpr[U,\psibar,\psi] = S[U,\psibar,\psi]
    +\delta S_{\rm v}[U,\psibar,\psi]
    +\delta S_{\rm G,b}[U]
    +\delta S_{\rm F,b}[U,\psibar,\psi],
\end{equation}
has the same structure as in the standard framework, in particular,
$\delta S_{\rm v}$ and $\delta S_{\rm G,b}$ are as given in eqs.~(I.5.3)
and~(I.5.6). The symmetries allow for two new fermionic 
boundary counterterms,   
\begin{equation}
  {\cal O}_\pm = i\muq\psibar\gamma_5\tau^3 P_\pm\psi.
  \label{counter}
\end{equation}
The equations of motion do not lead to a further reduction and
the action with the fermionic boundary counterterms at O($a$) is
then given by 
\begin{eqnarray}
  \delta S_{\rm F,b}[U,\psibar,\psi]&=&
  a^4\sum_{\bf x}\Bigl\{
  (\ctildes-1)\bigl[\ophat{s}({\bf x})+\ophatprime{s}({\bf x})\bigr]
  \nonumber\\
  &&\hphantom{0123} 
  +(\ctildet-1)\bigl[\ophat{t}({\bf x})-\ophatprime{t}({\bf x})\bigr]
  \nonumber\\[1ex] 
  &&\hphantom{0123} 
   +(\tilde{b}_1-1)\bigl[\widehat{Q}_1({\bf x})+\widehat{Q}_1'({\bf x})\bigr]
  \nonumber\\[1ex]
  &&\hphantom{0123} 
   +(\tilde{b}_2-1)\bigl[\widehat{Q}_2({\bf x})+\widehat{Q}_2'({\bf x})\bigr]\Bigr\}.
 \end{eqnarray}
Here, we have chosen lattice operators as follows,
\begin{eqnarray}
  \widehat{Q}_1({\bf x}) &=& 
             i\muq\psibar(x)\gamma_5\tau^3\psi(x)\bigl\vert_{x_0=a},\\[1ex]
  \widehat{Q}'_1({\bf x})&=& 
             i\muq\psibar(x)\gamma_5\tau^3\psi(x)\bigl\vert_{x_0=T-a},\\[1ex]
  \widehat{Q}_2({\bf x}) &=& i\muq\rhobar({\bf x})\gamma_5\tau^3\rho({\bf x}),\\[1ex]
  \widehat{Q}'_2({\bf x})&=& 
    i\muq\rhobarprime({\bf x})\gamma_5\tau^3\rhoprime({\bf x}),
\end{eqnarray}
and the expressions for the lattice operators 
$\ophat{s,t}$ and $\ophatprime{s,t}\,\,$
are given in eqs.~(I.5.21)--(I.5.24). Note that the improvement
coefficients are the same for both boundaries, as the counterterms
are related by a time reflection combined with a flavour exchange.

\subsection{Dirac equation and classical solutions}

For Euclidean times $0<x_0<T$ the lattice Dirac operator 
and its adjoint are formally defined through
\begin{eqnarray}
  {\delta \Simpr\over\vphantom{\vbox{\vskip2.3ex}}\delta\psibar(x)}
   &=& (D+\delta D+m_0+i\muq\gamma_5\tau^3)\psi(x),\\
 -{\delta \Simpr\over\delta\psi(x)} &=&
  \psibar(x)(\lvec{D}^{\dagger}+\delta\lvec{D}^{\dagger}+m_0+i\muq\gamma_5\tau^3),
\end{eqnarray}
where $\delta D=\delta D_{\rm v}+\delta D_{\rm b}$ is the sum of
the volume and the boundary O($a$) counterterms.
Eq.~(II.2.3) for the volume counterterms remains valid, whereas
for the boundary counterterms one obtains
\begin{eqnarray}
  \delta D_{\rm b}\psi(x)&=&
  (\ctildet-1){1\over a}\Bigl\{
  \delta_{x_0,a}
  \left[\psi(x)-U(x-a\hat{0},0)^{-1}P_{+}\psi(x-a\hat{0})\right]\nonumber\\
  &&\hphantom{01234567}+\delta_{x_0,T-a}
  \left[\psi(x)-U(x,0)P_{-}\psi(x+a\hat{0})\right] \Bigr\}\nonumber\\
  &&\mbox{}+(\tilde{b}_1-1) 
  \left[\delta_{x_0,a}+\delta_{x_0,T-a}\right]i\muq\gamma_5\tau^3\psi(x).
\end{eqnarray}
We observe that the net effect of the additional counterterm 
consists in the  replacement  $\muq\rightarrow\tilde{b}_1\muq$ close
to the boundaries. Although a boundary O($a$) effect is unlikely
to have a major impact, we note that the presence of this
counterterm with a general coefficient $\tilde{b}_1$
invalidates the argument by which zero modes of the Wilson-Dirac
operator are absent in twisted mass lattice QCD.
To circumvent this problem we remark that the counterterm 
may also be implemented by explicit insertions 
into the correlation functions. As every insertion
comes with a power of $a$, a single insertion will be
sufficient in most cases, yielding a result that is
equivalent up to terms of O($a^2$).

Given the Dirac operator, the propagator is now defined through
\begin{equation}
  (D+\delta D+m_0+i\muq\gamma_5\tau^3)S(x,y)=a^{-4}\delta_{xy}, \qquad
  0<x_0<T,
  \label{propagator}
\end{equation}
and the boundary conditions 
\begin{equation}
  P_{+}S(x,y)|_{x_0=0}=P_{-}S(x,y)|_{x_0=T}=0.
\end{equation}
Boundary conditions in the second argument follow from
the conjugation property,
\begin{equation}
  S(x,y)^{\dagger}=\gamma_5\tau^1 S(y,x)\gamma_5\tau^1,
  \label{conjugate}
\end{equation}
which is the usual one up to an exchange of the flavour components.

As in the standard framework~\cite{StefanII,paperII}, it is useful
to consider the classical solutions of the Dirac equation, 
\begin{eqnarray}
  \Bigl(D+\delta D+m_0+i\muq\gamma_5\tau^3\Bigr)\psiclass(x)&=& 0, \\
  \psibarclass(x)\left(\lvec{D}^{\dagger}+\delta\lvec{D}^{\dagger}
    +m_0+i\muq\gamma_5\tau^3\right)&=&0.
\end{eqnarray}
Here, the time argument is restricted to $0<x_0<T$,
while at the boundaries the classical solutions are
required to satisfy the inhomogeneous boundary conditions~(\ref{bcs}).
It is not difficult to obtain the explicit expressions,
\begin{eqnarray}
  \psiclass(x) &=& \ctildet a^3\sum_{\bfy}
  \Bigl\{S(x,y)U(y-a\hat{0},0)^{-1}P_+\rho(\bfy)\bigl\vert_{y_0=a}\nonumber\\ 
  &&\hphantom{\ctildet a^3\sum_{\bfy}}   
  +S(x,y)U(y,0)P_-\rhoprime(\bfy)\bigl\vert_{y_0=T-a}\Bigr\},
  \label{classical1}\\[1ex]
  \psibarclass(x)&=&\ctildet a^3\sum_{\bfy}
  \Bigl\{\rhobar(\bfy) P_-U(y-a\hat{0},0)S(y,x)\bigl\vert_{y_0=a}\nonumber\\ 
  &&\hphantom{\ctildet a^3\sum_{\bfy}}   
  +\rhobarprime(\bfy)P_+U(y,0)^{-1}S(y,x)\bigl\vert_{y_0=T-a}\Bigr\},
  \label{classical2}
\end{eqnarray}
which are again valid for $0<x_0<T$. Note that these
expressions are exactly the same as in ref.~\cite{paperII},
except that the quark propagator here is the 
solution of eq.~(\ref{propagator}).

\subsection{Quark functional integral and basic 2-point functions}

We shall use the same formalism for the quark functional integral
as described in subsect.~II.2.3. Most of the equations can 
be taken over literally, in particular, eq.~(II.2.21) holds again.
The presence of the twisted mass term merely leads to a modification
of the improved action of the classical fields, [eq.~(II.2.22)], which
is now given by
\begin{eqnarray}
  S_{\rm F,impr}[U,\psibarclass,\psiclass] &=& a^3\sum_{\bfx}\Bigl\{
   \tilde{b}_2a\muq \bigl[\rhobar(\bfx)i\gamma_5\tau^3\rho(\bfx)
                    +\rhobarprime(\bfx)i\gamma_5\tau^3\rhoprime(\bfx)\bigr]
    \nonumber\\[1ex]
  &&\hphantom{1234}
  +\ctildes a\bigl[\rhobar(\bfx)\gamma_k\frac12(\nab{k}+\nabstar{k})\rho(\bfx)
    \nonumber\\[1ex]
  &&\hphantom{12+\ctildes a\bigl[\rhobar}
  +\rhobarprime(\bfx)\gamma_k\frac12(\nab{k}+\nabstar{k})\rhoprime(\bfx)\bigr]
    \nonumber\\[1ex]
  &&\hphantom{1234}
  -\ctildet\bigl[\rhobar(\bfx)U(x-a\hat0,0)\psiclass(x)\bigl\vert_{x_0=a}
    \nonumber\\[1ex]
   &&\hphantom{a^3\sum_{\bfx}12}
   +\rhobarprime(\bfx)U(x,0)^{-1}\psiclass(x)\bigr\vert_{x_0=T-a}\bigr]\Bigr\}.
\end{eqnarray}
The quark action is a quadratic form in the Grassmann fields,
and the functional integral can be solved explicitly. Therefore, in a
fixed gauge field background any fermionic correlation function
can be expressed in terms of the basic two-point functions. 
Besides the propagator already introduced above,
\begin{equation}
  \left[\psi(x)\psibar(y)\right]_{\rm F} = S(x,y),
\end{equation}
we note that the boundary-to-volume correlators  
can be written in a convenient way
using the classical solutions,
\begin{eqnarray}
  \left[\zeta(\bfx)\psibar(y)\right]_{\rm F} &=&
     {\delta\psibarclass(y)\over{\delta\rhobar(\bfx)}},\\
  \left[\psi(x)\zetabar(\bfy)\right]_{\rm F} &=& 
     {\delta\psiclass(x)\over{\delta\rho(\bfy)}},\\
  \left[\zetaprime(\bfx)\psibar(y)\right]_{\rm F} &=&
     {\delta\psibarclass(y)\over{\delta\rhobarprime(\bfx)}},\\
  \left[\psi(x)\zetabarprime(\bfy)\right]_{\rm F} &=& 
     {\delta\psiclass(x)\over{\delta\rhoprime(\bfy)}}.
\end{eqnarray}
The explicit expressions in terms of the quark propagator
can be easily obtained from eqs.~(\ref{classical1})
and (\ref{classical2}), and
coincide with those given in ref.~\cite{paperII}. The
boundary-to-boundary correlators can be written 
as follows,
\begin{eqnarray}
  \left[\zeta({\bf x})\zetabarprime({\bf y})\right]_{\rm F}&=&
  \ctildet  P_-U(x-a\hat0,0)\left[\psi(x)
  \zetabarprime(\bfy)\right]_{\rm F}\Bigr|_{x_0=a},\\[1ex]
  \left[\zetaprime({\bf x})\zetabar({\bf y})\right]_{\rm F}&=&
     \ctildet P_+U(x,0)^{-1}
   \left[\psi(x)\zetabar(\bfy)\right]_{\rm F}\Bigr|_{x_0=T-a}.
\end{eqnarray}
The correlators of two boundary quark fields at the same boundary
receive additional contributions due to the new boundary counterterms, viz.
\begin{eqnarray}
  \left[\zeta({\bf x})\zetabar({\bf y})\right]_{\rm F}&=&
  \ctildet^2
  \left.P_{-}U(x-a\hat{0},0)S(x,y)U(y-a\hat{0},0)^{-1}P_{+}\right|_{x_0=y_0=a}
  \nonumber\\[1ex]
  &&\mbox{}-P_{-}\left[\ctildes\dirac{k}\frac12(\nabstar{k}+\nab{k})
  +\tilde{b}_2i\muq\gamma_5\tau^3\right]a^{-2}\delta_{\bf xy},\\[1ex]
  \left[\zetaprime({\bf x})\zetabarprime({\bf y})\right]_{\rm F}&=&
  \ctildet^2
  \left.P_{+}U(x,0)^{-1}S(x,y)U(y,0)P_{-}\right|_{x_0=y_0=T-a}\nonumber\\[1ex]
  &&\mbox{}-P_{+}\left[\ctildes\dirac{k}\frac12(\nabstar{k}+\nab{k})
  +\tilde{b}_2i\muq\gamma_5\tau^3\right]a^{-2}\delta_{\bf xy}.
\end{eqnarray}
We finally note that the conjugation property~(\ref{conjugate})
implies,
\begin{eqnarray}
  \left[\psi(x)\zetabar(\bfy)\right]_{\rm F}^\dagger &=&
  \gamma_5\tau^1\left[\zeta(\bfy)\psibar(x)\right]_{\rm
  F}\gamma_5\tau^1,
\\[1ex]
  \left[\zeta({\bf x})\zetabarprime({\bf y})\right]_{\rm F}^\dagger &=&
  \gamma_5\tau^1
  \left[\zetaprime({\bf y})\zetabar({\bf x})\right]_{\rm
  F}\gamma_5\tau^1,
\\[1ex]
  \left[\zeta({\bf x})\zetabar({\bf y})\right]_{\rm F}^\dagger &=&
  \gamma_5\tau^1
  \left[\zeta({\bf y})\zetabar({\bf x})\right]_{\rm F}\gamma_5\tau^1,
\end{eqnarray}
and analogous equations for the remaining 2-point functions.

\subsection{SF Correlation functions}

With this set-up of the SF we now define a few on-shell 
correlation functions involving the composite fields of Sect.~2.
With the boundary source
\begin{equation}
  {\cal O}_{}^{a} = a^6\sum_{\bfy,\bfz} 
  \zetabar(\bfy)\gamma_5\frac12 \tau^a\zeta({\bfz}),
\end{equation}
we define the correlation functions 
\begin{eqnarray}
  f_{\rm A}^{ab}(x_0)&=& -\langle A_0^a(x){\cal O}_{}^b\rangle,\\[1ex]
  f_{\rm P}^{ab}(x_0)&=& -\langle P^a(x){\cal O}_{}^b\rangle,\\[1ex]
  f_{\rm V}^{ab}(x_0)&=& -\langle V_0^a(x){\cal O}_{}^b\rangle.
\end{eqnarray}
In the following  we restrict the isospin indices to $a,b\in\{1,2\}$.
It is convenient to define the matrix~\cite{paperIII,paperIV}, 
\begin{equation}
  H(x) = a^3\sum_{\bfy} {\delta\psiclass(x)\over{\delta\rho({\bf y})}}.
\end{equation}
Its  hermitian conjugate matrix is given by
\begin{equation}
H(x)^\dagger = a^3\sum_{\bfy} \gamma_5\tau^1
  {\delta\psibarclass(x)\over{\delta\rhobar(\bfy)}}\gamma_5\tau^1,
 \end{equation}
and the correlation functions can be expressed in terms
of $H(x)$, viz.
\begin{equation}
  f_{\rm X}^{ab}(x_0)= \left\langle
        \frac14\tr\left\{H(x)^\dagger\gamma_5
   \Gamma_{\rm X}\tau^1\tau^a H(x)\tau^b\tau^1\right\}
                        \right\rangle_{\rm G}.
  \label{fx}
\end{equation}
As in ref.~\cite{paperII} the bracket $\langle\cdots\rangle_{\rm G}$ 
means an average over the gauge fields with the effective
gauge action,
\begin{equation}
  \Seff[U]= \Sg[U]+\delta S_{\rm G,b}[U]
  -\ln\det\left(D+\delta D+m_0+i\muq\gamma_5\tau^3\right),
  \label{Seff}
\end{equation}
and the trace is over flavour, Dirac and colour indices.
The gamma structures are  $\Gamma_{\rm X}=\gamma_0\gamma_5,\gamma_5,\gamma_0$, 
where  ${\rm X}$ stands for ${\rm A},{\rm P}$ and ${\rm V}$ respectively.

\subsection{Reducing the flavour structure}

In order to  carry out the flavour traces we introduce
the flavour projectors 
\begin{equation}
  Q_\pm=\frac12(1\pm\tau^3).
\end{equation}
Inserting the flavour decomposition, 
\begin{equation}
  H(x)=  H_+(x)Q_+ +  H_-(x)Q_-,
\end{equation}
into the expression eq.~(\ref{fx}) leads to 
\begin{equation}
  \fx^{ab}(x_0) = \sum_{i,j=\pm}\tr\{Q_i\tau^1\tau^aQ_j\tau^b\tau^1\}
                        \left\langle
        \frac14\tr\left\{H_i(x)^\dagger\gamma_5
            \Gamma_{\rm X} H_j(x)\right\}
                        \right\rangle_{\rm G}.
\end{equation}
Since we restrict the indices $a$ and $b$ to values in $\{1,2\}$ 
this expression further simplifies leading to
\begin{equation}
  \fx^{ab}(x_0) = \sum_{i=\pm}\tr\{Q_i\tau^b\tau^a\}
                        \left\langle
        \frac14\tr\left\{H_i(x)^\dagger\gamma_5\Gamma_{\rm X} H_i(x)\right\}
                        \right\rangle_{\rm G}.
  \label{fxab}
\end{equation}
In order to simplify the expressions further, we now 
study the behaviour under a parity transformation combined
with the exchange $\muq \rightarrow -\muq$. 
Notice that the parity transformation also transforms the
background fields, in particular it implies $\theta_k\rightarrow
-\theta_k$ $(k=1,2,3)$.
On the matrices $H_\pm(x)$ this transformation acts according to
\begin{equation}
   H_\pm(x)\longrightarrow \gamma_0 H_\mp(\tilde{x}),
\end{equation}
where  $\tilde{x}=(x_0,-\bfx)$ is the parity transformed space-time
argument, and we recall that $H_\pm(x)$ depend implicitly on the
background gauge field. After averaging over the gauge
fields and due to parity invariance of the effective
gauge action~(\ref{Seff}) one then finds
\begin{equation}
  \left\langle
    \tr\left\{H_\pm(x)^\dagger
  \gamma_5\Gamma_{\rm X} H_\pm(x)\right\}
  \right\rangle_{\rm G} = 
   \eta({\rm X})\left\langle\tr\left\{H_\mp(x)^\dagger
  \gamma_5\Gamma_{\rm X} H_\mp(x)\right\}
  \right\rangle_{\rm G}, 
\label{symmetry}
\end{equation}
where the sign factor depends on whether
$\Gamma_{\rm X}$ commutes ($\eta({\rm X})=-1$) or anti-commutes 
($\eta({\rm X})=1$) with $\gamma_0$.
Using this result in eq.~(\ref{fxab}) it follows that 
\begin{equation}
   \fa^{12}(x_0)=\fp^{12}(x_0)=\fv^{11}(x_0)=0.
\end{equation}
Furthermore, the exact U(1) flavour symmetry implies that
\begin{equation}
 \fx^{22}(x_0)=\fx^{11}(x_0),\qquad \fx^{21}(x_0)=-\fx^{12}(x_0),
\end{equation}
so that we may restrict attention to the 
following non-vanishing correlation functions:
\begin{eqnarray}
  \fa^{11}(x_0)&=& -\frac12 
                   \left\langle
                   \tr\left\{
                     H_+(x)^\dagger\gamma_0 H_+(x)
                     \right\}
                   \right\rangle_G,\\[1ex]
  \fp^{11}(x_0)&=&  \frac12 
                   \left\langle
                    \tr \left\{
                     H_+(x)^\dagger H_+(x)
                       \right\}
                   \right\rangle_G,\\[1ex]
  \fv^{12}(x_0)&=& \frac{i}2 
                   \left\langle
                   \tr \left\{
                     H_+(x)^\dagger\gamma_0\gamma_5 H_+(x)
                       \right\}
                   \right\rangle_G.
\end{eqnarray}
Note that eq.~(\ref{symmetry}) has allowed to eliminate the
dependence upon the second flavour component  $H_-(x)$. 
This is convenient both for perturbative calculations
and in the framework of numerical simulations.

\section{O($a$) improvement of the free theory}

We determine the improvement coefficients in the free theory,
which is obtained by setting all gauge links to unity.
In this context correlation functions of quark and
antiquark fields are suitable on-shell quantities
which ought to be improved.  We may therefore 
consider the improvement of the one-particle energies, 
the quark propagator and basic 2-point functions 
in the Schr\"odinger functional, in addition to 
the SF correlation functions introduced in section~3. 

\subsection{The free quark propagator}

All correlation functions in the SF are obtainable from
the quark propagator, which can be computed
using standard methods~\cite{paperII}.
We set the standard improvement coefficients
to their known values~\cite{paperII},
\begin{equation}
   \ctildet=\ctildes=1, 
  \label{treeimprov}
\end{equation}
and compute  the propagator assuming $\tilde{b}_1=1$. 
As discussed in sect.~3, any other value can be
obtained by insertion of the corresponding boundary counterterm.
The propagator can be written in the form
\begin{equation}
   S(x,y)=\left(D^\dagger+m_0-i\muq\gamma_5\tau^3\right)G(x,y),
  \label{qprop}
\end{equation}
where $G(x,y)$ is given by
\begin{equation}
  G(x,y)=L^{-3}\sum_{\bfp}\rme^{i\bfp(\bfx-\bfy)}
  \left[G_+(\bfp, x_0,y_0)P_+ +G_-(\bfp, x_0,y_0)P_-\right],
\end{equation}
with the functions
\begin{eqnarray}
 G_+(\bfp; x_0, y_0) &=& 
 {\cal N}(\momp{p}{})  \biggl\{   
  M_-(\momp{p}{})\left[\rme^{-\omega({\bf\smomp{p}{}})(|x_0-y_0|-T)}
        -\rme^{\omega({\bf\smomp{p}{}})(x_0+y_0-T)}\right] \nonumber\\[1ex]
 &&\mbox{}\!\!\!\!\!\! 
  +M_+(\momp{p}{})\left[ \rme^{\omega({\bf\smomp{p}{}})(|x_0-y_0|-T)}
        -\rme^{-\omega({\bf\smomp{p}{}})(x_0+y_0-T)}\right]\biggr\}, \\[1ex]
  G_-(\bfp; x_0, y_0) &=&  G_+(\bfp; T-x_0, T-y_0).
\end{eqnarray}
Here, $M_\pm(\momp{p}{})=M(\momp{p}{})\pm i\momp{\ring{p}}{0}$~(II.3.17),
with $M(p)$  as defined in eq.~(II.3.6) and $p_\mu^+=p_\mu+\theta_\mu/L$.
Furthermore, we recall that in the above formulae
it is understood that $p_0=p_0^+=i\omega({\bf\smomp{p}{}})$, where
for given spatial momentum $\bfq$ the energy $\omega(\bfq)$ is obtained as
the solution of the equation
\begin{equation}
  \sinh\left[{a\over2}\,\omega({\bf q})\right]=
  {a\over2}
  \left\{
  {{\bf\ring{q}}^2+\muq^2+(m_0+\frac{1}{2}a{\bf\hat{q}}^2)^2
  \over
  1+a(m_0+\frac{1}{2}a{\bf\hat{q}}^2)}
  \right\}^{\frac{1}{2}}.
  \label{omega}
\end{equation}
Finally, using again the notation of ref.~\cite{paperII},
the normalization factor is given by
\begin{equation}
  {\cal N}(\momp{p}{})=
  \left\{-2i\momp{\ring{p}}{0}A({\bf\momp{p}{}})R(\momp{p}{})
  \rme^{\omega({\bf\smomp{p}{}})T}\right\}^{-1}.
\end{equation}

\subsection{Improvement conditions and results}

In the free quark theory, the quark energy $\omega$ is a suitable 
on-shell quantity. At zero spatial momentum it
coincides with the pole mass, which is related to 
the bare masses through
\begin{equation}
  \cosh a\mpole = 1+\frac12 a^2(m_0^2+\muq^2)/(1+am_0).
\end{equation}
Up to terms of O($a^2$) one then finds ($\mc=0$ at tree level)
\begin{equation}
  \mpole^2=\left(\mq^2+\muq^2\right)\left(1-a\mq\right)+\rmO(a^2).
\end{equation}
Replacing the bare masses by the renormalized O($a$) improved
mass parameters and requiring the absence of O($a$) 
artifacts one obtains
\begin{equation}
  \bm=-\frac12,\qquad \bmu+\tbm+\frac12=0,
  \label{mpole_imp}
\end{equation}
and the same condition is obtained from the O($a$) improved
energy at finite spatial momentum.
One may wonder whether it is possible to get an additional
condition by considering the improvement of the quark
propagator itself. This is not so, for the reasons given 
in subsection~2.5. As an illustration we consider the
quark propagator~(\ref{qprop}) in the limit of infinite
time extent $T$ with the limit taken at fixed 
$x_0-T/2$ and $y_0-T/2$.
This eliminates the boundaries both at $x_0=0$ and $x_0=T$,
so that one is left with the improvement of the mass parameters,
and of the quark and antiquark fields, viz.
\begin{eqnarray}
  \psi_{\rm R} &=& 
  \left(1+b_\psi am_0 + \tilde{b}_\psi ia\muq\gamma_5\tau^3\right)\psi, \\
  \psibar_{\rm R} &=& 
 \psibar\left(1+b_{\psibar} am_0 + \tilde{b}_{\psibar} 
 ia\muq\gamma_5\tau^3\right).
\end{eqnarray}
Requiring the quark propagator to 
be O($a$) improved we find the usual result of
the untwisted theory, $b_\psi=b_{\psibar}=\frac12$, and 
\begin{equation}
 \tilde{b}_{\psibar}=\tilde{b}_\psi,\qquad
  2\tilde{b}_\psi-\tbm-\frac12=0,\qquad 
 2\tilde{b}_\psi+ b_\mu=0,
\end{equation}
i.e.~3 equations for 4 coefficients. Similarly, 
by studying the SF correlation functions of the improved
quark bilinear fields we find the standard results
of the untwisted theory, $\ca=\cv=0$ and $2\,b_\zeta=\ba=\bv=\bp=1$,
and the following conditions involving the new coefficients, 
\begin{eqnarray}
   \tilde{b}_1-\frac12(\tbm+\frac12) &=& 1,\\
  \bmu+\tbm+\frac12                  &=& 0,\\
   \tba -(\tbm+\frac12)                &=& 0,\\
   \tbv -(\tbm+\frac12)                &=& 0.
\end{eqnarray}
Furthermore, from the O(a) improvement of the basic 2-point functions
we also obtain
\begin{equation}
   \tilde{b}_2=1.
\end{equation}
The fact that $\tilde{b}_\psi$ and $\tilde{b}_1$ are not determined
independently is again due to the invariance of the continuum 
theory under axial rotations of the fields and a compensating
change in the mass parameters. Hence our findings in
the free theory are completely in line with
the general expectation expressed in subsect.~2.5.
Choosing $\tbm$ as the free parameter and setting it to $-\frac12$ leads
to $\bmu=\tba=\tbv=0$ and $\tilde{b}_1=1$, while e.g. for $\tbm=0$
the tree level value $\tilde{b}_1=5/4$ is somewhat inconvenient.

\section{The one-loop computation}

We now want to expand the correlation functions to one-loop order.
We work with vanishing  boundary values $C_k^{}$ and $C_k'$.
The gauge fixing procedure then is the same as
in ref.~\cite{paperII} and will not be described here.
In the following we only describe those aspects that are
new and otherwise assume the reader to be familiar with
refs.~\cite{paperII,PeterStefan}.

\subsection{Renormalized amplitudes}

Once the flavour traces have been taken, the one-loop calculation 
at fixed lattice size is almost identical 
to the standard case~\cite{paperII,PeterStefan}.
In order to take the continuum limit at fixed physical space-time
volume, we then keep $\mr$, $\mur$, $x_0$ and  $T$
fixed in units of $L$. 
Here the renormalized mass parameters are defined in
a mass-independent renormalization scheme which may remain unspecified
for the moment.

To first order of perturbation theory
the substitutions for the coupling constant and the quark mass 
then amount to
\begin{eqnarray} 
   g_0^2 &=& \gr^2+\rmO(\gr^4),\\
   m_0   &=& m_0^{(0)}+\gr^2 m_0^{(1)}+\rmO(\gr^4),\\
   \muq  &=& \muq^{(0)}+\gr^2 \muq^{(1)}+\rmO(\gr^4),
\end{eqnarray}
where the precise form of the coefficients
\begin{eqnarray} 
  m_0^{(0)}  &=& {1\over a}\Bigl[1-\sqrt{1-2a\mr-a^2\mur^2}\Bigr],
  \label{m00}\\[1ex]
  m_0^{(1)}  &=& \mc^{(1)}-\biggl\{
        \zm^{(1)}\mr+\bm^{(1)}a\left(m_0^{(0)}\right)^2\nonumber\\
  &&\mbox{}\hphantom{}   
  +a\mur^2\left[\tbm^{(1)}+\Zmu^{(1)}+\bmu^{(1)}am_0^{(0)}\right]\biggr\}
   \left[1-am_0^{(0)}\right]^{-1},
  \label{m01}\\[1ex]
  \muq^{(0)} &=& \mur,  
  \label{muq0}\\[1ex]
  \muq^{(1)} &=& -\muq^{(0)}\left\{Z_\mu^{(1)}+\bmu^{(1)}am_0^{(0)}\right\},
  \label{muq1} 
\end{eqnarray}
is a direct consequence of the definitions made in subsect.~2.1,
and already includes the tree-level results obtained in the
preceding section with the particular choice $\tbm^{(0)}=-\frac12$.

The renormalized correlation functions,
\begin{eqnarray}
  [\fv^{12}(x_0)]_{{\hbox{\sixrm R}}}&=&
  \zv(1+\bv a\mq)\zzeta^2(1+\bzeta a\mq)^2\nonumber\\[1ex]
  &&\times
   \left\{\fv^{12}(x_0)+\tbv a\muq\fa^{11}(x_0)\right\},\\[1ex]
   {[}\fp^{11}(x_0)]_{{\hbox{\sixrm R}}}&=&
  \zp(1+\bp a\mq)\zzeta^2(1+\bzeta a\mq)^2 \fp^{11}(x_0),\\[1ex]
   {[}\fa^{11}(x_0)]_{{\hbox{\sixrm R}}}&=&
  \za(1+\ba a\mq)\zzeta^2(1+\bzeta a\mq)^2\nonumber\\[1ex]
  &&\times \left\{\fa^{11}(x_0)+\ca
  a\tilde{\partial}_0\fp^{11}(x_0)-\tba a\muq\fv^{12}(x_0)\right\}, 
\end{eqnarray}
have a well-defined perturbation expansion in the renormalized
coupling $\gr$, with coefficients that are computable 
functions of $a/L$. 
For instance the expansion of $[\fv^{12}]_{{\hbox{\sixrm R}}}$ reads
\begin{eqnarray}
  [\fv^{12}(x_0)]_{{\hbox{\sixrm R}}}&=&
  \fv^{12}(x_0)^{(0)}+\gr^2\Bigl\{\fv^{12}(x_0)^{(1)}
  +m_0^{(1)}{\partial\over\partial m_0}\fv^{12}(x_0)^{(0)}\nonumber\\
  &&\quad+\left(\zv^{(1)}+2\zzeta^{(1)}
    +a\mr\bigl[\bv^{(1)}+2\bzeta^{(1)}\bigr]\right)\fv^{12}(x_0)^{(0)}\nonumber\\
  &&\mbox{}+\muq^{(1)}{\partial\over\partial \muq}\fv^{12}(x_0)^{(0)}
     +a\mur\tbv^{(1)}\fa^{11}(x_0)^{(0)}\Bigr\},
\end{eqnarray}
where terms of order $a^2$ and $\gr^4$ have been neglected, and
it is understood that the correlation functions are evaluated
at $m_0=m_0^{(0)}$ and $\muq=\muq^{(0)}$.

Following ref.~\cite{paperII} we now set  $x_0=T/2$ and 
scale all dimensionful quantities in units of $L$. 
With the parameters $z_m=\mr L$, $z_\mu=\mur L$
and  $\tau=T/L$ we then consider the dimensionless functions,
\begin{eqnarray}
  \ha(\theta,z_m,z_\mu,\tau,a/L)&=&
  [\fa^{11}(x_0)]_{{\hbox{\sixrm R}}}\bigr|_{x_0=T/2},\\[1ex]
  \hv(\theta,z_m,z_\mu,\tau,a/L)&=& 
  [\fv^{12}(x_0)]_{{\hbox{\sixrm R}}}\bigr|_{x_0=T/2},\\[1ex]
  \hp(\theta,z_m,z_\mu,\tau,a/L)&=&
  [\fp^{11}(x_0)]_{{\hbox{\sixrm R}}}\bigr|_{x_0=T/2},\\[1ex]
  \hda(\theta,z_m,z_\mu,\tau,a/L)&=&
  L\tilde\partial_0
  [\fa^{11}(x_0)]_{{\hbox{\sixrm R}}}\bigr|_{x_0=T/2},\\[1ex]
  \hdv(\theta,z_m,z_\mu,\tau,a/L)&=&
  L\tilde\partial_0
  [\fv^{12}(x_0)]_{{\hbox{\sixrm R}}}\bigr|_{x_0=T/2}.
\end{eqnarray}
One then infers,
\begin{eqnarray}
  \ha &=& v_0+\gr^2\biggl\{v_1+\ctildet^{(1)}v_2+a m_0^{(1)}v_3+
  \ca^{(1)}v_4+a\muq^{(1)}v_5 \nonumber\\ 
   &&\mbox{}\hphantom{0123456}
   +z_\mu\tilde{b}_1^{(1)}v_6 -{a\over L}z_\mu\tba^{(1)}q_0\nonumber\\
  &&\mbox{}\hphantom{0123456}
   +\left(\za^{(1)}+2\zzeta^{(1)}+
   {{a}\over{L}}z_m\bigl[\ba^{(1)}+2\bzeta^{(1)}\bigr]\right)v_0
   \biggr\}, 
   \label{ha1loop}\\[1ex]
  \hv &=& q_0+\gr^2\biggl\{q_1+\ctildet^{(1)}q_2+a m_0^{(1)}q_3+
   a\muq^{(1)}q_5 \nonumber\\
   &&\mbox{}\hphantom{0123456}  
   +z_\mu\tilde{b}_1^{(1)}q_6 +{a\over L}z_\mu\tbv^{(1)}v_0\nonumber\\
   &&\mbox{}\hphantom{0123456}
   +\left(\zv^{(1)}+2\zzeta^{(1)}+
   {{a}\over{L}}z_m\bigl[\bv^{(1)}+2\bzeta^{(1)}\bigr]\right)q_0
   \biggr\}, 
   \label{hv1loop}\\[1ex]
  \hp &=& u_0+\gr^2\biggl\{u_1+\ctildet^{(1)}u_2+a m_0^{(1)}u_3+
  a\muq^{(1)}u_5 +z_\mu \tilde{b}_1^{(1)}u_6\nonumber\\
  &&\mbox{}\hphantom{0123456}
  +\left(\zp^{(1)}+2\zzeta^{(1)}+
  {{a}\over{L}}z_m\bigl[\bp^{(1)}+2\bzeta^{(1)}\bigr]\right)u_0
  \biggr\}, 
  \label{hp1loop}\\[1ex]
  \hda &=& w_0+\gr^2\biggl\{w_1+\ctildet^{(1)}w_2+a m_0^{(1)}w_3+
  \ca^{(1)}w_4+a\muq^{(1)}w_5 \nonumber\\ 
  &&\mbox{}\hphantom{0123456}
  +z_\mu\tilde{b}_1^{(1)}w_6 -{a\over L}z_\mu\tba^{(1)}r_0\nonumber\\
  &&\mbox{}\hphantom{0123456}
  +\left(\za^{(1)}+2\zzeta^{(1)}+
  {{a}\over{L}}z_m\bigl[\ba^{(1)}+2\bzeta^{(1)}\bigr]\right)w_0
  \biggr\},
  \label{hda1loop}\\[1ex]
  \hdv &=& r_0+\gr^2\biggl\{r_1+\ctildet^{(1)}r_2+a m_0^{(1)}r_3+
  a\muq^{(1)}r_5 \nonumber\\
  &&\mbox{}\hphantom{0123456}
  +z_\mu\tilde{b}_1^{(1)}r_6 +{a\over L}z_\mu\tbv^{(1)}w_0\nonumber\\
  &&\mbox{}\hphantom{0123456}
  +\left(\zv^{(1)}+2\zzeta^{(1)}+
  {{a}\over{L}}z_m\bigl[\bv^{(1)}+2\bzeta^{(1)}\bigr]\right)r_0
  \biggr\}.
  \label{hdv1loop}
\end{eqnarray}
Since we are neglecting terms of order $a^2$, the expansions,
\begin{eqnarray}
 m_0^{(1)} &=& \mc^{(1)}-\Zm^{(1)}{z_m\over L}-
  {az_m^2\over L^2}\left[\Zm^{(1)}+\bm^{(1)}\right]
 -{az_\mu^2\over L^2}\left[Z_\mu^{(1)}+\tbm^{(1)}\right],\\[1ex]
 \muq^{(1)} &=& -{z_\mu\over L} 
  \left[Z_\mu^{(1)}+\bmu^{(1)}{az_m\over L}\right]
\end{eqnarray}
may be inserted in Eqs.~(\ref{ha1loop})-(\ref{hdv1loop}). 
All the coefficients $v_i,\dots,r_i$ are still functions 
of $\tau,\,,\theta\,,z_m$ and $z_\mu$. 
Analytic expressions can be derived
for those coefficients involving the tree level
correlation functions or the O($a$) counterterms. 
Their asymptotic expansions for  $a/L\rightarrow 0$ 
are collected in Appendix~B.
The coefficients $v_1,\dots,r_1$ are only obtained numerically and
definite choices for the parameters had to be made.
We generated numerical data for $\theta=0$ and $\theta=0.5$
for both $T=L$ and $T=2L$ and various combinations of 
the mass parameters $z_m$ and $z_\mu\ne0$ with values between 0 and
1.5. With these parameter choices the Feynman diagrams 
were then evaluated numerically in 64 bit precision arithmetic for
a sequence of lattice sizes ranging from $L/a=4$ to $L/a=32$ 
(and in some cases to $L/a=36$).

\subsection{Analysis and results}

The renormalization constants are
determined by requiring the renormalized amplitudes
to be finite in the continuum limit, and by the requirement
that the tmQCD Ward identities be satisfied~\cite{tmQCD1}.
A linear divergence is cancelled in all amplitudes
by inserting the usual one-loop coefficient $am_c^{(1)}$,
or equivalently a series which converges to this coefficient
in the limit $a/L\rightarrow 0$~\cite{paperII}.
We choose the  lattice minimal-subtraction scheme
to renormalize the pseudo-scalar density and the quark
boundary fields, and the one-loop coefficients are
then given by [with $\CF=(N^2-1)/2N$],
\begin{equation}
  \Zp^{(1)}=-{6\cf\over 16\pi^2}\ln (L/a)\,, \qquad
  2\Zz^{(1)}=-\Zp^{(1)}\,.
  \label{Zequalities}
\end{equation}
The current renormalization constants,
and the renormalization of the standard and twisted
mass parameters are determined by the Ward identities.
For the one-loop coefficients 
we expect~\cite{GabrielliEtAl,Stefan96,tmQCD1},
\begin{eqnarray}
 \Za^{(1)}  &=&   -0.087344(2)\,\cf, \\[1ex]
 \Zv^{(1)}  &=&   -0.097072(2)\,\cf,  \\[1ex]
 \Zm^{(1)}  &=& -\Zp^{(1)} - 0.019458(1)\,\cf, \\[1ex]
 \Zmu^{(1)} &=& -\Zp^{(1)}\,. 
\end{eqnarray}
With our data we were able to compute the one-loop coefficients
of the combinations $\Zm\Zp/\Za$ and $Z_\mu\Zp/\Zv$, 
as well as the logarithmically divergent parts of all 
one-loop coefficients. Complete consistency with the above
expectations was found, and we shall adopt these results in the following.

The corresponding coefficients in other
schemes differ from those above by $a$-independent terms.
With the renormalization constants chosen in this way we 
find e.g. for the combination of separately diverging terms appearing
in the curly bracket of (\ref{hp1loop})   
\begin{eqnarray}
  u_1+a\mc^{(1)}u_3&+&(\Zp^{(1)}+2\Zz^{(1)})u_0
  -\Zm^{(1)} z_mu_3^{(-1)}-\Zmu^{(1)} z_\mu u_5^{(-1)}
  \nonumber\\[1ex]
  &=&{\cal U}_0+{\cal U}_1{a\over L}+\rmO(a^2/L^2)\,,
  \label{ufinite}
\end{eqnarray}
where ${\cal U}_i$ are functions of $\tau\,,\theta\,,z_m$ and $z_\mu$, and 
$u_i^{(-1)}$ are coefficients of $L/a$ in the expansion of $u_i$
for $L/a\rightarrow \infty$. Evidently similar equations 
hold for the other functions $v_1,q_1,w_1,r_1$.
It is important to note that we expect no terms
involving $(a/L)\ln(L/a)$ on the right hand side of (\ref{ufinite})
because we have imposed tree level improvement, and this
was indeed seen in our data analysis.  
Moreover there are no terms  $\sim\Zm^{(1)}a/L$ or
$\sim\Zmu^{(1)}a/L$ on the left hand side above because 
of Eq.~(\ref{u3m1}); thus the coefficient ${\cal U}_1$ is 
(contrary to ${\cal U}_0$) independent of the 
renormalization scheme. Estimates for the coefficients 
${\cal U}_1$,${\cal V}_1,\dots$ were obtained for the
various data sequences using the methods described in~\cite{extrapoln}. 

Now the improvement coefficients are determined by demanding
that the renormalized amplitudes approach the continuum limit 
with corrections of O($a^2/L^2$). For the cancellation of
the O$(a)$ terms the following equations should be satisfied (for 
undefined notation see Appendix~B):  
\begin{eqnarray}
  z_\mu\Bigl[\,z_\mu\tbm^{(1)}v_3^{(-1)}+\,\,z_m\bmu^{(1)}v_5^{(-1)}
  +\tba^{(1)}q_0^{(0)}
  -\tilde{b}_1^{(1)}v_6^{(1)}\Bigr]
  &=& {\cal V}_1+\bar{{\cal V}}_1\,,\label{vequation} \\[1ex]
  z_\mu\Bigl[z_\mu\tbm^{(1)}q_3^{(-1)}+\,\,z_m\bmu^{(1)}q_5^{(-1)}
  -\tbv^{(1)}v_0^{(0)}-\tilde{b}_1^{(1)}q_6^{(1)}\Bigr]
  &=& {\cal Q}_1+\bar{{\cal Q}}_1\,, \label{qequation} \\[1ex]
  z_\mu\Bigl[z_\mu\tbm^{(1)}u_3^{(-1)}+\,\,z_m\bmu^{(1)}u_5^{(-1)} 
  \phantom{-\tba^{(1)}r_0^{(0)}}\, -\tilde{b}_1^{(1)}u_6^{(1)}\Bigr]
  &=& {\cal U}_1+\bar{{\cal U}}_1\,, \label{uequation} \\[1ex]
  z_\mu\Bigl[z_\mu\tbm^{(1)}w_3^{(-1)}+z_m\bmu^{(1)}w_5^{(-1)}
  +\tba^{(1)}r_0^{(0)} -\tilde{b}_1^{(1)}w_6^{(1)}\Bigr]
  &=& {\cal W}_1+\bar{{\cal W}}_1\,, \label{wequation} \\[1ex]
  z_\mu\Bigl[\,z_\mu\tbm^{(1)}r_3^{(-1)}\,+\,z_m\bmu^{(1)}r_5^{(-1)}
  -\tbv^{(1)}w_0^{(0)} -\tilde{b}_1^{(1)}r_6^{(1)} \Bigr]
  &=& {\cal R}_1+\bar{{\cal R}}_1\,. \label{requation}
\end{eqnarray}
In these equations all terms involving improvement 
coefficients which are necessary also in the untwisted 
theory, have been collected in the terms $\bar{{\cal U}}_1,\dots$
on the right hand sides and they are specified in equations
(\ref{bvequation})-(\ref{brequation}). 
The numerical values of these improvement coefficients, 
obtained in previous analyses ~\cite{paperII,PeterStefan},
are:
\begin{eqnarray}
\ctildet^{(1)}&=&-0.01346(1)\,\CF\,,
  \\[1ex]
\ca^{(1)}&=&-0.005680(2)\,\CF\,, 
  \\[1ex]
\bzeta^{(1)}&=&-0.06738(4)\,\CF\,, 
  \\[1ex]
\bm^{(1)}&=&-0.07217(2)\,\CF\,, 
  \\[1ex]
\ba^{(1)}&=&\phantom{-}0.11414(4)\,\CF\,, 
 \\[1ex]
\bv^{(1)}&=&\phantom{-}0.11492(4)\,\CF\,, 
  \\[1ex]
\bp^{(1)}&=&\phantom{-}0.11484(4)\,\CF\,. 
\end{eqnarray}

Before we proceed with the numerical analysis of
equations (\ref{vequation})-(\ref{requation}),
it is essential to note that
using the identities (\ref{uidentity})-(\ref{ridentity})
they can be rewritten as
\begin{eqnarray}
  z_\mu\Bigl[z_m\bmu^{\prime(1)}v_5^{(-1)}
  +\tba^{\prime(1)}q_0^{(0)}
  -\tilde{b}_1^{\prime(1)}v_6^{(1)}\Bigr]
  &=& {\cal V}_1+\bar{{\cal V}}_1\,,\label{vequationp} \\[1ex]
  z_\mu\Bigl[z_m\bmu^{\prime(1)}q_5^{(-1)}
  -\tbv^{\prime(1)}v_0^{(0)}
  -\tilde{b}_1^{\prime(1)}q_6^{(1)}\Bigr]
  &=& {\cal Q}_1+\bar{{\cal Q}}_1\,, \label{qequationp} \\[1ex]
  z_\mu\Bigl[z_m\bmu^{\prime(1)}u_5^{(-1)} 
  \phantom{-\tba^{\prime(1)}r_0^{(0)}}\,
  -\tilde{b}_1^{\prime(1)}u_6^{(1)}\Bigr]
  &=& {\cal U}_1+\bar{{\cal U}}_1\,, \label{uequationp} \\[1ex]
  z_\mu\Bigl[z_m\bmu^{\prime(1)}w_5^{(-1)}
  +\tba^{\prime(1)}r_0^{(0)} 
  -\tilde{b}_1^{\prime(1)}w_6^{(1)}\Bigr]
  &=& {\cal W}_1+\bar{{\cal W}}_1\,, \label{wequationp} \\[1ex]
  z_\mu\Bigl[z_m\bmu^{\prime(1)}r_5^{(-1)}
  -\tbv^{\prime(1)}w_0^{(0)} 
  -\tilde{b}_1^{\prime(1)}r_6^{(1)} \Bigr]
  &=& {\cal R}_1+\bar{{\cal R}}_1\,, \label{requationp}
\end{eqnarray}
where the primed coefficients appearing here are defined through
\begin{eqnarray}
\bmu^{\prime(1)}&=&\bmu^{(1)}+\tbm^{(1)}\,,
\\
\tilde{b}_1^{\prime(1)}&=&\tilde{b}_1^{(1)}-\frac12\tbm^{(1)}\,,
\\
\tba^{\prime(1)}&=&\tba^{(1)}-\tbm^{(1)}\,,
\\
\tbv^{\prime(1)}&=&\tbv^{(1)}-\tbm^{(1)}\,.
\end{eqnarray}
In other words, from our equations we can only obtain information
on four linearly independent combinations of the new improvement
coefficients appearing in the twisted theory.
This was in fact to be anticipated from our general
discussion in subsect.~2.5, where we argued that we are free to chose
for example the coefficient $\tbm^{(1)}$ as we please. 

Since our equations are over-determined and 
also having generated such a large selection of data sets, 
we had many ways to proceed to determine the
coefficients $\bmu^{\prime(1)}, \tilde{b}_1^{\prime(1)}, \tba^{\prime(1)}$
and $\tbv^{\prime(1)}$,
and a multitude of consistency checks on the results.
We first note that if we consider the linear combination of amplitudes 
$\hda-2z_m\hp$ and $\hdv+2z_\mu\hp$ associated with the
PCAC and PCVC relations respectively we obtain
\begin{eqnarray}
  -2z_\mu^2u_0^{(0)}\tba^{\prime(1)}
  &=&{\cal W}_1+\bar{{\cal W}}_1
  -2z_m\left({\cal U}_1+\bar{{\cal U}}_1\right)\,,\label{pcacrel}\\[1ex]
  -2z_\mu z_m u_0^{(0)}\left(\bmu^{\prime(1)}+\tbv^{\prime(1)}\right)  
  &=&{\cal R}_1+\bar{{\cal R}}_1
  +2z_\mu\left({\cal U}_1+\bar{{\cal U}}_1\right)\,.\label{pcvcrel}
\end{eqnarray}
With knowledge of the right hand sides, each equation
determines a particular linear combination of improvement coefficients.
In these equations the boundary coefficient
$\tilde{b}_1^{(1)}$ does not appear as expected.
On the other hand the coefficient $\tilde{b}_1^{\prime(1)}$
is all that appears on the left hand sides of 
Eqs.~(\ref{uequationp}),(\ref{requationp}) for the data sets with $z_m=0$.

By solving simultaneously the three equations (\ref{vequationp}),
(\ref{uequationp}) and (\ref{wequationp}) 
for one data set with $z_m\ne0$, we could obtain the three 
coefficients\footnote{Particularly good results were obtained 
e.g.~with the data set $z_m=0\,,z_\mu=0.5\,,\theta=0$, where we in 
fact had data up to $L/a=36$.}
$\bmu^{\prime(1)}, \tba^{\prime(1)}$ and $\tilde{b}_1^{\prime(1)}$
(and of course analogously for the equations involving the vector
current). We also extracted
the two coefficients $\bmu^{\prime(1)}, \tilde{b}_1^{\prime(1)}$   
by solving just Eq.~(\ref{uequationp}) for two different data sets (of
which at least one has $z_m\ne0$).

Unfortunately due to rounding errors, 
the one-loop cutoff effects like ${\cal U}_1$
were rarely determined better than to within a few percent.
The consequence of this was that many routes of analyses 
described above and when applied to various (combinations of)
data sets, led to results for the improvement coefficients
with very large errors.
Nevertheless there remained sufficiently many analyses which
delivered useful results with relatively small errors,
and in these cases all results were consistent with each other
and with our following ``best estimates":
\begin{eqnarray}
\bmu^{\prime(1)}&=&-0.103(3)\,\CF\,,
\\
\tilde{b}_1^{\prime(1)}&=&\phantom{-}0.035(2)\,\CF\,,
\\
\tba^{\prime(1)}&=&\phantom{-}0.086(4)\,\CF\,,
\\
\tbv^{\prime(1)}&=&\phantom{-}0.074(3)\,\CF\,.
\end{eqnarray}

As one practical choice for applications in numerical simulations
we advocate $\tbm=-\frac12$ to all orders of perturbation theory,
which would result in setting $\tbm^{(1)}=0$ in the above equations.

\section{Conclusions}

In this paper we have introduced the set-up of O($a$) improved
twisted mass lattice QCD in its simplest form with two mass-degenerate
quarks. In perturbation theory to one-loop order we have 
verified that O($a$) improvement works out as expected. 
We have identified the new counterterms and computed their
coefficients at the tree-level and to one-loop order.
In practice perturbative estimates may be satisfactory, 
as tmQCD has been primarily designed to explore the chiral region of QCD,
where the contribution of the new counterterms should be small anyway. 
This expectation is confirmed by a non-perturbative scaling
test in a physically small volume, which employs the 
perturbative values of the new improvement coefficients
reported here~\cite{tmQCDscaling}. However, a non-perturbative
determination of some of the new coefficients is certainly desirable and
may be possible along the lines of ref.~\cite{babp}.

An interesting aspect of O($a$) improved tmQCD is the absence
of any new counterterm corresponding to a rescaling of 
the bare coupling $g_0$. This singles out the choice 
for the angle $\alpha=\pi/2$ for which the physical quark mass
is entirely defined in terms of the twisted mass parameter.
A quark mass dependent rescaling of $g_0$ is hence
completely avoided, and one may hope 
that this eases the chiral extrapolation or interpolation 
of numerical simulation data. Furthermore, 
using the over-completeness of the counterterms 
(cf. subsect.~2.5) to fix $\tbm$ exactly,
no tuning is necessary to obtain $\alpha=\pi/2$ up to O($a^2$) effects, 
provided the standard critical mass $\mc$ and the standard improvement
coefficients of the massless theory $\csw$ and $\ca$ are known. 
We also note that, at $\alpha=\pi/2$, both sides of the exact PCVC relation 
are automatically renormalized and O($a$) improved. 
This can be exploited for an O($a$) improved determination
of $F_\pi$~\cite{Milano}, as the vector current at $\alpha=\pi/2$ is
physically interpreted as the axial current~\cite{tmQCD1}.

In the future one may wish to extend the framework 
of O($a$) improved tmQCD to include
the heavier quarks in the way suggested in ref.~\cite{tmQCD1}.
The analysis of O($a$) counterterms still remains to be done,
but we do not expect any new conceptual problems here.

Finally, we have defined the Schr\"odinger functional for tmQCD,
based on the appropriate generalisation of L\"uscher's 
transfer matrix construction for tmQCD. We expect that 
the Schr\"odinger functional will be useful in 
the determination of hadronic matrix elements along the
lines of refs.~\cite{SFscaling,SFhadron}, and work in this direction
is currently in progress~\cite{Milano,B_K}.

\vskip 1ex

This work is part of the ALPHA collaboration research programme.
We are grateful to P.A.~Grassi for discussions and 
his collaboration in the tmQCD project. Thanks also go to M.~L\"uscher,
R.~Sommer and  A.~Vladikas for useful comments and discussions.  
S.~Sint acknowledges partial support by the European Commission 
under grant No.~FMBICT972442.

\appendix
\renewcommand{\thesection}{A}
\section{The transfer matrix for twisted mass lattice QCD}
\renewcommand{\thesection}{A}
In this appendix we briefly indicate the generalization
of the transfer matrix construction for twisted mass
lattice QCD with $\csw=0$. 
We use the original notation of ref.~\cite{transfer} with the 
conventions of ref.~\cite{StefanI}.
The transfer matrix as an operator in Fock space and as an
integral kernel with respect to the gauge fields
has the structure
\begin{equation}
  {T}_0[U,U'] = \hat T_{\rm F}^\dagger(U)K_0[U,U']\hat T_{\rm F}^{}(U'),
\label{transfer}
\end{equation}
with pure gauge kernel $K_0$ and the fermionic part
\begin{equation}
   \hat T_{\rm F}(U)=\det(2\kappa B)^{1/4}
   \exp(\hat{\chi}^\dagger P_-C\hat\chi)
   \exp(-\hat{\chi}^\dagger\gamma_0 M\hat\chi).        
\end{equation}
Here, the operators $\hat{\chi}_i({\bf x})$ are canonical 
($i$ is a shorthand for colour, spin and flavour indices) 
viz. 
\begin{equation}
   \{\hat\chi_i^{}(\bfx),\hat\chi_j^\dagger(\bfy)\}
   =\delta_{ij}a^{-3}\delta_{\bfx\bfy},  
\end{equation}
and $B$ and  $C$ are matrix representations of the
difference operators
\begin{eqnarray}
 B &=& 1 - 6\kappa -a^2\kappa \sum_{k=1}^3\nabla_k^\ast\nabla_k, \\
 C &=& a\sum_{k=1}^3 \gamma_k\frac12(\nab{k}+\nabstar{k})  
            +ia\muq\gamma_5\tau^3.
\end{eqnarray}
As in the standard case the positivity of the transfer matrix 
hinges on the positivity of  the matrix $B$, which is 
guaranteed for $||\kappa||<1/6$. This is the standard 
bound which also ensures that the matrix $M$,
\begin{equation}
   M = \frac12 \ln\left(\frac12 B\kappa^{-1}\right),
\end{equation}
is well-defined. No restriction applies 
to the twisted mass parameter, except that
$\muq$ must be real for the transfer matrix~(\ref{transfer}) to reproduce
the twisted mass lattice QCD action.

\appendix
\renewcommand{\thesection}{B}
\section{Analytic expressions for expansion coefficients}
\renewcommand{\thesection}{B}
In this appendix we provide explicit analytic expressions
for the tree-level amplitudes and the counterterms appearing in
eqs.~(\ref{ha1loop})--(\ref{hdv1loop})
which are needed to compute the one-loop amplitudes up to terms of O($a^2$).
We have checked that the analytic expressions correctly
reproduce the numerical values obtained by directly programming
the correlation functions and counterterm insertions.

First we define 
\begin{eqnarray}
  \omega &=& \sqrt{z_m^2+3\theta^2+z_\mu^2}\,,\\
  \co     &=& \cosh(\omega \tau)\,,\\
  \si     &=& \sinh(\omega \tau)\,,\\
  \rho    &=& \omega  \co+z_m \si\,,\\
  \nu     &=& \omega  \si+z_m \co\,,
\end{eqnarray}
where $\tau=T/L$. Then we have $u_0=u_0^{(0)}+\rmO(a^2/L^2)$ etc.~with 
\begin{eqnarray}
  u_0^{(0)} &=& {N\omega \over \rho}\,,\\
  v_0^{(0)} &=& -{N(3\theta^2+z_\mu^2+z_m\nu)\over \rho^2}\,,\\
  q_0^{(0)} &=& {Nz_\mu(-z_m+\nu)\over \rho^2}\,,\\
  w_0^{(0)} &=& 2z_m u_0^{(0)}\,,\\
  r_0^{(0)} &=& -2z_\mu u_0^{(0)}\,.
\end{eqnarray}
For the boundary terms we define
\begin{equation}
  \hat{r}=z_m+{2(3\theta^2+z_\mu^2)\si\over\rho}\,,
\end{equation}
and then $u_2=au_2^{(1)}/L+\rmO(a^2/L^2)$ etc.~with
\begin{eqnarray}
  u_2^{(1)} &=& 2\hat{r}u_0^{(0)}\,,\\
  v_2^{(1)} &=& 2\hat{r}v_0^{(0)}
                -{4N\omega (3\theta^2+z_\mu^2)(-z_m+\nu)\over\rho^3}\,,\\
  q_2^{(1)} &=& 2\hat{r}q_0^{(0)}
                -{4N\omega  z_\mu(3\theta^2+z_\mu^2+z_m\nu)\over\rho^3}\,,\\
  w_2^{(1)} &=& 2\hat{r}w_0^{(0)}\,,\\
  r_2^{(1)} &=& 2\hat{r}r_0^{(0)}\,.  
\end{eqnarray}
Similarly, $u_6=au_6^{(1)}/L+\rmO(a^2/L^2)$ etc.~with
\begin{eqnarray}
  u_6^{(1)}&=&-{2z_\mu\si\over\rho}u_0^{(0)}\,,\\
  v_6^{(1)}&=&-{2z_\mu\si\over\rho}v_0^{(0)}
             +{2N\omega  z_\mu(-z_m+\nu)\over\rho^3}\,,\\
  q_6^{(1)}&=&-{2z_\mu\si\over\rho}q_0^{(0)}
             +{2N\omega \left(\omega\rho+z_\mu^2(1-\co)\right)\over\rho^3}\,,\\
  w_6^{(1)}&=&-{2z_\mu\si\over\rho}w_0^{(0)}\,,\\
  r_6^{(1)}&=&-{2z_\mu\si\over\rho}r_0^{(0)}\,.
\end{eqnarray}
For the derivatives with respect to the mass parameters we have,
\begin{equation}
  u_i=(L/a)u_i^{(-1)}+u_i^{(0)}+\rmO(a/L), \quad (i=3,5)
\end{equation}
with
\begin{equation}
  u_3^{(0)}=-z_m u_3^{(-1)},\quad u_5^{(0)}=-z_\mu u_3^{(-1)},
 \label{u3m1}
\end{equation}
and analogous equations hold in all other cases.
Defining
\begin{eqnarray}
  X         &=& {\nu(1+z_m\tau)\over\omega }\,, \\
  Y         &=& {\rho(1+z_m\tau)\over\omega }\,,\\
  \tilde{X} &=& {z_\mu(\nu\tau+\co)\over\omega }\,,\\
  \tilde{Y} &=& {z_\mu(\rho\tau+\si)\over\omega }\,,
\end{eqnarray}
one has
\begin{eqnarray}
  u_3^{(-1)} &=& -{Xu_0^{(0)}\over \rho}
                     +{Nz_m\over \omega \rho}\,,\\
  v_3^{(-1)} &=& -{2Xv_0^{(0)}\over \rho}
                 -{N(\nu+z_m Y)\over \rho^2}\,,\\
  q_3^{(-1)} &=& -{2Xq_0^{(0)}\over \rho}
                 -{Nz_\mu(1-Y)\over \rho^2}\,,\\
  w_3^{(-1)} &=& 2(z_m u_3^{(-1)}+u_0^{(0)})\,,\\
  r_3^{(-1)} &=& -2z_\mu u_3^{(-1)}\,,
\end{eqnarray}
and
\begin{eqnarray}
  u_5^{(-1)} &=& -{\tilde{X}u_0^{(0)}\over \rho}
                     +{Nz_\mu\over \omega \rho}\,,\\
  v_5^{(-1)} &=& -{2\tilde{X}v_0^{(0)}\over\rho}
                 -{N(2z_\mu+z_m\tilde{Y})\over \rho^2}\,,\\
  q_5^{(-1)} &=& -{2\tilde{X}q_0^{(0)}\over\rho}
                 +{N(z_\mu\tilde{Y}-z_m+\nu)\over \rho^2}\,,\\
  w_5^{(-1)} &=& 2z_mu_5^{(-1)}\,,\\
  r_5^{(-1)} &=& -2z_\mu u_5^{(-1)}-2u_0^{(0)}\,.
\end{eqnarray}
Note the identities
\begin{eqnarray}
0 &=& 2z_\mu u_3^{(-1)}-2z_m u_5^{(-1)}\,-u_6^{(1)}\,,
\label{uidentity}
\\
0 &=& 2z_\mu v_3^{(-1)}-2z_m v_5^{(-1)}\,-v_6^{(1)}+2q_0^{(0)}\,,
\label{videntity}
\\
0 &=& 2z_\mu q_3^{(-1)}-2z_m q_5^{(-1)}\,\,-q_6^{(1)}-2v_0^{(0)}\,,
\label{qidentity}
\\
0 &=& 2z_\mu w_3^{(-1)}-2z_m w_5^{(-1)}-w_6^{(1)}+2r_0^{(0)}\,,
\label{widentity}
\\
0 &=& 2z_\mu r_3^{(-1)}-2z_m r_5^{(-1)}\,\,-r_6^{(1)}-2w_0^{(0)}\,.
\label{ridentity}
\end{eqnarray}
The remaining coefficients to be specified 
are $v_4=av_4^{(1)}/L+\rmO(a^2/L^2)$ and
$w_4=aw_4^{(1)}/L+\rmO(a^2/L^2)$ with
\begin{eqnarray}
  v_4^{(1)}&=&-{2N\omega ^2\nu\over\rho^2}\,,\\
  w_4^{(1)}&=&{4N\omega^3\over\rho}\,.
\end{eqnarray}

Finally we specify the terms $\bar{\cal U}_1,\dots$ appearing on
the right hand side of equations (\ref{vequation})--(\ref{requation}):
\begin{eqnarray}
  \bar{{\cal V}}_1&=&
  \ctildet^{(1)}v_2^{(1)}
  -z_m^2\bm^{(1)}v_3^{(-1)}
  +z_m[\ba^{(1)}+2\bzeta^{(1)}]v_0^{(0)}+\Ca^{(1)}v_4\,,\label{bvequation}
\\[1ex]
  \bar{{\cal Q}}_1&=&
  \ctildet^{(1)}q_2^{(1)}
  -z_m^2\bm^{(1)}q_3^{(-1)}
  +z_m[\bv^{(1)}+2\bzeta^{(1)}]q_0^{(0)}\,,
  \label{bqequation}
\\[1ex]
  \bar{{\cal U}}_1&=&
  \ctildet^{(1)}u_2^{(1)}
  -z_m^2\bm^{(1)}u_3^{(-1)}
  +z_m[\bp^{(1)}+2\bzeta^{(1)}]u_0^{(0)}\,,
  \label{buequation}
\\[1ex]
  \bar{{\cal W}}_1&=&
  \ctildet^{(1)}w_2^{(1)}
  -z_m^2\bm^{(1)}w_3^{(-1)}
  +z_m[\ba^{(1)}+2\bzeta^{(1)}]w_0^{(0)}+\Ca^{(1)}w_4\,,
  \label{bwequation}
\\[1ex]
  \bar{{\cal R}}_1&=&
  \ctildet^{(1)}r_2^{(1)}
  -z_m^2\bm^{(1)}r_3^{(-1)}
  +z_m[\bv^{(1)}+2\bzeta^{(1)}]r_0^{(0)}\,.
  \label{brequation}
\end{eqnarray}


\end{document}